\pdfoutput=1
\documentclass[aps,prl,twocolumn,superscriptaddress,longbibliography]{revtex4-2}
\usepackage{graphicx}
\usepackage{amsmath,amssymb,bm}
\usepackage{xcolor}
\usepackage{hyperref}
\hypersetup{colorlinks=true,linkcolor=blue!55!black,citecolor=blue!55!black,urlcolor=blue!55!black}
\graphicspath{{figs/}}

\begin{document}

\title{Swapping Quantum Annealing Errors into a Cavity}

\author{Hao Zhang}
\email{hao.zhang.quantum@gmail.com}
\affiliation{Department of Physics, University of Wisconsin--Madison, Madison, Wisconsin 53706, USA}

\begin{abstract}
Quantum annealing is a promising form of quantum computation, but it
slows down at a small energy gap $\Delta$ during the anneal, and the
errors it makes there remain undetected. We introduce cavity quantum
annealing, in which a cavity mode of frequency $O(1)\gg\Delta$, coupled
mid-schedule, swaps these errors into photons. In the $p$-spin model,
this reduces the annealing time to nearly the square root of that
required without the cavity. Moreover, counting the photons certifies the
result: high-photon runs, with an $O(1)$ yield, reach a ground-state fidelity above $99.99\%$ orders of
magnitude sooner, a gain that grows exponentially with system size. A pre-loaded photon instead runs the swap in reverse,
preparing an excited state. More broadly, our work shows that cavities
can enhance computations performed with qubits, paving the way toward
high-fidelity quantum state preparation by quantum annealing.
\end{abstract}

\maketitle

\begin{figure}[t]
\includegraphics[width=\columnwidth]{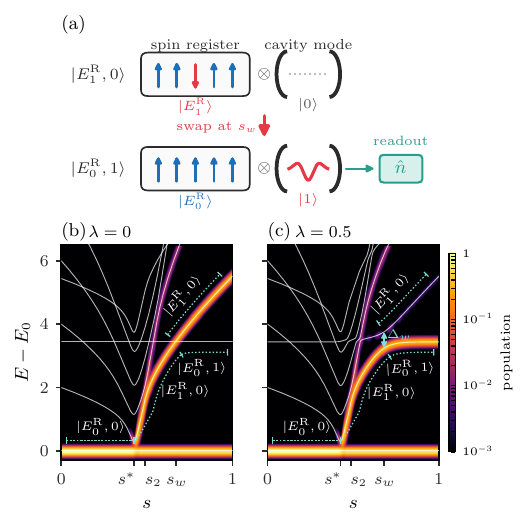}
\caption{Swapping errors into a cavity. (a)~The swap exchanges
$|E^{\rm R}_1,0\rangle$, the register in its first excited level (the
error, red) with the cavity empty, and $|E^{\rm R}_0,1\rangle$, the
register in its ground level with one photon, at $s_w$; the readout
$\hat n$ counts photons.
(b,c)~Population of the instantaneous levels (log scale) during an
anneal from $s=0$ to $1$, for $T=169$,
$N=24$, $\omega_c=3.44$; $s^*$ marks the minimum gap, $s_2$ the second
crossing, and $s_w$ the swap crossing. (b)~Without coupling
($\lambda=0$) the population excited at $s^*$ stays on
$|E^{\rm R}_1,0\rangle$, passing straight through its crossing with
$|E^{\rm R}_0,1\rangle$ at $s_w$: the error remains in the register. (c)~With coupling ($\lambda=0.5$) the two
levels repel at $s_w$ (gap $\Delta_w$), and the excited population
follows $|E^{\rm R}_0,1\rangle$: the error becomes a photon.}
\label{fig1}
\end{figure}

\begin{figure*}[t]
\includegraphics[width=\textwidth]{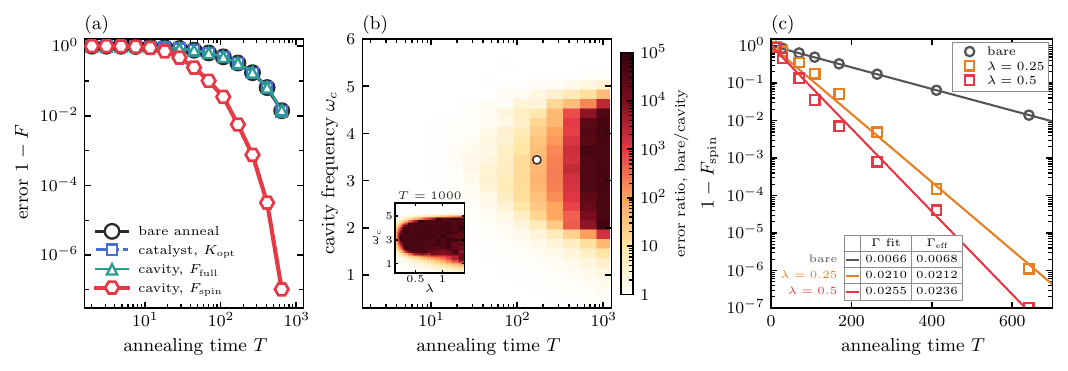}
\caption{The cavity makes the anneal faster, while a catalyst
acting on the spins does not. Data for $N=24$. (a)~Error $1-F$ versus annealing time. The bare anneal, the
catalyst at its optimal strength $K_{\rm opt}$, and the cavity
($F_{\rm full}$) coincide; the cavity ($F_{\rm spin}$) lies several orders of magnitude below.
The cavity uses $\lambda=0.5$ and the best $\omega_c$ at each $T$.
(b)~Ratio of the bare to the cavity ($F_{\rm spin}$) error at
$\lambda=0.5$; the circle marks the Fig.~\ref{fig1} run; inset, the same ratio at $T=1000$ over $(\omega_c,\lambda)$. (c)~Error
versus $T$ at $\omega_c=3.44$ for two couplings,
with one-parameter fits $e^{-\Gamma T}$ (lines); the table compares the fitted rates with
Eq.~\eqref{eq:geff}.}
\label{fig2}
\end{figure*}

\emph{Introduction.---}Quantum annealing prepares the ground state of
a problem Hamiltonian by sweeping slowly from a simple initial
Hamiltonian \cite{Finnila1994,Kadowaki1998,Brooke1999,Farhi2001,%
Santoro2002,Das2008,AlbashLidar2018}. Over the past decades it has
produced many promising results: adiabatic state preparation \cite{Farhi2000,AspuruGuzik2005,Aharonov2007,%
Islam2011,Bernien2017,Ebadi2021,Scholl2021,Semeghini2021,%
GonzalezCuadra2025,Mu2025,Bombieri2025,Yue2026,Zeng2026}, combinatorial
optimization and sampling
\cite{Johnson2011,Boixo2014,Harris2018,Hauke2020,Bauza2025,Camino2025,%
Li2026,Baldwin2026thr}, the
simulation of quantum dynamics and critical phenomena
\cite{King2018,King2022,King2023,King2025,Vodeb2025,Criticality2026,%
Teza2026,Pelofske2026}, and even the training of classical neural networks
\cite{Adachi2015,Amin2018,Laydevant2024,Xie2024,Zhang2025nn}. However, quantum
annealing faces a central obstruction, the energy gap. The adiabatic
theorem demands an
annealing time $T\gg\Delta^{-2}$, with $\Delta$ the minimum gap
between the ground and first excited states along the sweep
\cite{AlbashLidar2018}. For many problems the sweep meets a
first-order transition, where $\Delta$ closes exponentially with the
number of spins $N$ and the annealing time grows beyond reach. A
less discussed problem associated with the gap is that errors occur
silently. When the sweep is too fast the system crosses the gap into
an excited state and stays there, and nothing in the output says so.
If the ground state of the problem Hamiltonian is classical, a
product state, verifying the output amounts to computing its energy,
and for hard problems one cannot tell whether that energy is the
lowest; if it is quantum, a superposition, certifying it destroys it.
Thus the annealer returns an answer with no guarantee against error,
which we call the silent error problem.

The gap problem has been addressed in three ways, all aimed at
reaching the ground state. The sweep itself
can be reshaped, by optimized and inhomogeneous schedules and drivers
\cite{RolandCerf2002,Susa2018,Schloemer2025}, added spin interactions
(catalysts) or modified problem Hamiltonians that reshape the gap
\cite{SekiNishimori2012,Albash2019cat,Dadgar2026,Schulz2025}, counterdiabatic
driving \cite{Demirplak2003,Berry2009,Sels2017,Claeys2019,%
delCampo2013,Banks2025}, longitudinal bias fields and reverse
annealing \cite{Grass2019,Ohkuwa2018,Baldwin2026rev}, and iterative
cyclic annealing \cite{Wang2022mbl,Zhang2024cyclic}. The environment can also be
exploited. Thermal relaxation and pauses can assist
annealing \cite{Amin2008,Dickson2013,Marshall2019}, and engineered
dissipation can remove excitations from the qubits
\cite{Poyatos1996,Verstraete2009,Harrington2022,Mi2024,Torggler2017,%
Theis2018,ZhouWu2026,Polla2021,Ding2024,Zhan2026}. On the other hand, the
output can be post-processed after the anneal, as in digital cooling,
which removes excitations algorithmically from an ensemble of
annealing runs \cite{Zhang2025dc}.

These approaches avoid the error or discard it without a record,
leaving the silent error problem open. The error
itself, however, carries information: a run
whose error has been removed and counted brings with it evidence
that the spins are in the ground state. Here we propose cavity
quantum annealing, a protocol built on this observation. In the
protocol, one or more cavity modes \cite{Jarc2023,Sivak2023,Sauerwein2023,Young2024,Luo2025,Grinkemeyer2025,Keren2026,Helmrich2026}
are coupled to the annealer in the middle of the schedule. The error created at the minimum gap is
coherently swapped into a photon at a second level crossing created
by the cavity coupling, and the spins return to their ground state
(Fig.~\ref{fig1}). Despite its simplicity, the protocol shortens the
anneal substantially. With $T^{\rm cav}_{0.1}$ and $T^{\rm bare}_{0.1}$ the
annealing times that reach a fixed accuracy with and without the
cavity, we find roughly $T^{\rm cav}_{0.1}=(T^{\rm bare}_{0.1})^{1/2}$ across
system sizes, a square-root scaling speedup that can bring practical
annealers into the coherent regime \cite{King2022}. Central to the
protocol, the photons left in the cavity are the record of the
repair: we find that the more photons a run returns, the more likely
the spins are in their ground state, and the count is read without
disturbing the qubits, a certificate that conventional annealing
cannot provide. Beyond certification, a photon loaded before the
sweep runs the swap in reverse, an excited-state preparation beyond
current protocols.

\emph{Model and protocol.---}Cavity quantum annealing couples an
annealer of $N$ spins to cavity modes. We mainly discuss one mode and
return to two in the scaling analysis. The Hamiltonian $H(s)$ of the
coupled system, written as a function of the sweep parameter $s=t/T$
with $t$ the time and $T$ the total annealing time, is
\begin{equation}
\begin{split}
H(s) =\ & (1-s)H_D + sH_P + \omega_c a^\dagger a \\
        & + g(s)\,(a+a^\dagger)\,O + \frac{g(s)^2}{\omega_c}\,O^2 .
\end{split}
\label{eq:model}
\end{equation}
The first two terms are the conventional anneal, in which the driver $H_D$ is switched off and the problem Hamiltonian
$H_P$ switched on linearly. The third term is the energy of the
cavity field, a single mode of frequency $\omega_c$ with photon
annihilation and creation operators $a$ and $a^\dagger$ ($\hbar=1$). The last two
terms are the Pauli--Fierz dipole coupling and its dipole self-energy
\cite{RokajEtAl2018,SchaeferEtAl2020,RomanRoche2025}: the mode couples to a
collective spin operator $O$ through a pulse $g(s)=g_0 f(s)$ with
$f(s)=\sin^2(\pi s)$, which rises to a peak $g_0$ in the middle of
the sweep and vanishes at both ends, so the computational endpoints
are unmodified. To label the states during the sweep, we collect the terms acting
on the spins alone, the annealing terms and the self-energy, into a
spin-only Hamiltonian; we call the spins it governs the register, in
distinction to the cavity, and write
$H_R(s)=(1-s)H_D+sH_P+g(s)^2O^2/\omega_c$. Its instantaneous levels
$E^{\rm R}_k(s)$ and eigenstates $|E^{\rm R}_k(s)\rangle$, with $k=0$
the ground level, label the states below. The products
$|E^{\rm R}_k,n\rangle$ of $|E^{\rm R}_k(s)\rangle$ with $n$ photons serve
as a basis, and the instantaneous states of Eq.~\eqref{eq:model} are
labeled by their dominant projection onto it.

In this work we illustrate cavity quantum annealing on the
ferromagnetic $p$-spin model, the standard model of the first-order
gap problem \cite{Jorg2008,BapstSemerjian2012,SekiNishimori2012}. We
take $H_P=-N(2S_z/N)^p$ with $p=3$ and $H_D=-2S_x$, which also fixes
the unit of energy, with $S_x$ and $S_z$ the collective spin
operators $S_\alpha=\tfrac12\sum_i\sigma_i^\alpha$. Numerically,
the minimum gap of the conventional anneal $(1-s)H_D+sH_P$ closes as
$\Delta\approx e^{-0.06N}$ \cite{Jorg2008,BapstSemerjian2012}. The
cavity couples to the collective operator $O=2S_x/\sqrt N$
\cite{Dicke1954,TavisCummings1968}, which changes the minimum gap by
less than two percent. The initial state, the ground state of $H_D$
with all spins polarized along $x$, has maximal total spin $S=N/2$,
and by spin permutation symmetry, which $O$ preserves, the dynamics
stays in this $(N{+}1)$-dimensional sector. To describe the coupling across system sizes, we use the
dimensionless parameter $\lambda=g_0\sqrt N/\omega_c$ for the strength
of the spin--cavity coupling.

The fidelity $F$ of the final state $|\Psi\rangle$ at $s=1$ is scored in two ways,
\begin{equation}
F_{\rm full}=|\langle E^{\rm R}_0,0|\Psi\rangle|^2,\qquad
F_{\rm spin}=\sum_n|\langle E^{\rm R}_0,n|\Psi\rangle|^2,
\label{eq:scores}
\end{equation}
where $|E^{\rm R}_0\rangle$ is taken at $s=1$ and the sum runs over
the photon number $n$. The strict score $F_{\rm full}$ is the
fidelity with the ground state of the whole system at $s=1$, the
ground state of $H_P$ with zero photons, and is the fidelity that a
conventional annealing protocol reports. Since our target is the ground state of $H_P$ alone, the
practical score $F_{\rm spin}$ is the probability of finding the
register in that state, whatever the number of photons left in the
cavity. From this one can already anticipate a gain in performance,
since the ground state is reached through more channels, one for
each photon number.

\emph{Speedup.---}To understand where the speedup of cavity quantum
annealing comes from, and to distinguish it from mechanisms that act
on the spins alone, we compare three protocols: (A)~the
bare anneal, the conventional anneal without the cavity; (B)~a
catalyst, in which the last three terms of Eq.~\eqref{eq:model} are
replaced by $K\,[g(s)/g_0]^2 O^2$, with the strength $K$ scanned
over both signs and a range of magnitudes; this family covers the
static effect of a cavity that creates no real photons, namely the
dispersive shift from virtual photons, $K=-g_0^2/\omega_c$, and the
self-energy, $K=+g_0^2/\omega_c$; and (C)~cavity quantum annealing,
scanned over $(\omega_c,\lambda)$. All time evolutions are obtained by
exact numerical integration of the Schr\"odinger equation with
Eq.~\eqref{eq:model}, with the photon number truncated where the
results have converged.

We compare the three protocols at $N=24$ by the error $1-F$ of a single
run as a function of the annealing time [Fig.~\ref{fig2}(a)].
Measured against the bare anneal, cavity quantum annealing scored by
the spin readout lowers the error by several orders of magnitude at
fixed annealing time and shortens the annealing time to a fixed error
several-fold. This gain holds across a broad plateau in the system
parameters $(T,\omega_c,\lambda)$ [Fig.~\ref{fig2}(b)], so no careful
optimization is needed to reach it. The catalyst, even at its
optimal strength $K_{\rm opt}$, changes the fidelity by less than one
percent, and its curve collapses onto the bare one, as does that of
the cavity scored by $F_{\rm full}$. This agrees with the finding that
at $p=3$ the antiferromagnetic catalyst ($K>0$) cannot remove the
first-order transition \cite{SekiNishimori2012}; it can only steer
around the transition with precise prior knowledge of the location of
an $N$-dependent special path \cite{Durkin2019}. The cavity, in contrast, speeds
up the anneal for general $N$, crossing the first-order transition and
repairing the error by swapping it into a photon in the cavity
(Fig.~\ref{figscale} and below).

\emph{Swap mechanism.---}At finite $N$ the first-order transition
appears in the spectrum as an avoided crossing of minimum gap $\Delta$
at $s=s^*$ [Fig.~\ref{fig1}(b)]. An anneal of finite duration leaves
part of the population there in the first excited level,
$|E^{\rm R}_1,0\rangle$: this is the error. Past $s^*$ the error meets
a second crossing at $s=s_2$, which will be discussed later. The
gap $E^{\rm R}_1-E^{\rm R}_0$ then rises and, at $s=s_w$, sweeps
through the photon energy $\omega_c$, where $|E^{\rm R}_1,0\rangle$ meets
$|E^{\rm R}_0,1\rangle$ [Fig.~\ref{fig1}(c)]. At this swap crossing the
spin--photon coupling turns the error into a photon. The crossings at $s^*$ and $s_w$, both of Landau--Zener type
\cite{Landau1932,Zener1932}, are the main mechanism behind the final
fidelity. To calculate it, we follow the path that leaves the
error in the register: the run crosses $s^*$ diabatically, which
creates the error, and then crosses $s_w$ diabatically, which misses
the swap. The two crossing probabilities are \cite{Shevchenko2010}
\begin{equation}
P^{\rm cross}_{s^*}=e^{-\pi\Delta^2 T/(2v_1)},\qquad
P^{\rm cross}_{s_w}=e^{-\pi\Delta_w^2 T/(2v_w)}.
\label{eq:lz}
\end{equation}
Here $v_1$ is
the relative slope of $E^{\rm R}_0$ and $E^{\rm R}_1$ away from the
minimum gap;
$\Delta_w=2g(s_w)\,|\langle E^{\rm R}_1|O|E^{\rm R}_0\rangle|$ is the
gap of the swap crossing; and
$v_w=|d(E^{\rm R}_1{-}E^{\rm R}_0)/ds|_{s_w}$ is the rate at
which the register gap rises. Along this path the error is
the product of the two,
\begin{equation}
1-F_{\rm spin}=P^{\rm cross}_{s^*}\,P^{\rm cross}_{s_w}
= e^{-(\Gamma_1+\Gamma_w)T},
\label{eq:cascade}
\end{equation}
with $\Gamma_1=\pi\Delta^2/(2v_1)$ and
$\Gamma_w=\pi\Delta_w^2/(2v_w)$, the error decay rates of the two
crossings. Without the cavity the error decays at $\Gamma_1$ alone;
the cavity brings an additional decay, at rate $\Gamma_w$.

The additional decay is fast, $\Gamma_w\gg\Gamma_1$, because
$\Delta_w$ falls off algebraically with $N$ while $\Delta$ falls off
exponentially. The contrast comes from the branch structure of the
first-order transition. The problem has two macroscopically distinct
branches, a paramagnet with the spins along $x$ and a ferromagnet with
the spins along $z$, each with its own ladder of collective
excitations. At $s^*$ the ground state changes from the paramagnet to
the ferromagnet. States on opposite branches differ by an extensive
spin rearrangement and are connected only exponentially weakly, so
$\Delta$ is exponentially small. Past $s^*$ the paramagnetic level,
which carries the error, climbs through the ferromagnetic ladder, and
after it crosses $s_2$ the lowest levels are the ferromagnet and its
excitations. At $s_w$, $|E^{\rm R}_0\rangle$ and $|E^{\rm R}_1\rangle$
therefore lie on the \emph{same} branch and differ by one collective
excitation, so $\langle E^{\rm R}_1|O|E^{\rm R}_0\rangle=O(1)$. At
fixed $\lambda$ this leaves $\Delta_w\propto g_0=\lambda\omega_c/\sqrt N$,
which scales only as $N^{-1/2}$: a gap much larger than $\Delta$.

We now return to the second crossing at $s=s_2$, which opens a second
path that leaves the error in the register. There the paramagnetic
level meets the next ferromagnetic
level $|E^{\rm R}_2\rangle$ (Fig.~\ref{fig1}), with gap $\Delta_2$
and relative slope $v_2$. A run that crosses $s_2$ diabatically, with
probability $e^{-\Gamma_2T}$, where $\Gamma_2=\pi\Delta_2^2/(2v_2)$,
keeps the error beyond the reach of the swap. The error thus survives
along two paths,
\[
1-F_{\rm spin}=e^{-\Gamma_1T}\,\big(1-e^{-\Gamma_2T}\big)\,e^{-\Gamma_wT}
+e^{-\Gamma_1T}\,e^{-\Gamma_2T}.
\]
At long times the more slowly decaying term dominates, with effective
error decay rate
\begin{equation}
\Gamma_{\rm eff}=\Gamma_1+\min(\Gamma_w,\Gamma_2).
\label{eq:geff}
\end{equation}
Below a crossover coupling $\lambda^*$ the path through the swap
dominates and Eq.~\eqref{eq:cascade} holds. Above it the rate
saturates at $\Gamma_1+\Gamma_2$, and a stronger coupling brings no
further gain. The rate of Eq.~\eqref{eq:geff} agrees well with fits
to the numerical data [Fig.~\ref{fig2}(c)].

\begin{figure}[t]
\includegraphics[width=\columnwidth]{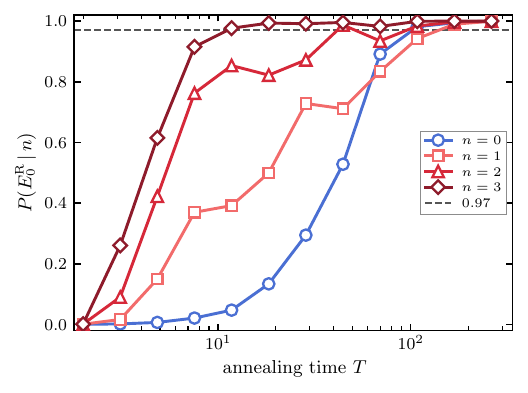}
\caption{The photon number certifies the ground state. Data for $N=24$,
$\omega_c=3.44$, $\lambda=0.5$. The reliability $P(E^{\rm R}_0|n)$, the
probability that the register is in the ground state given $n$
photons, grows with $n$, and three photons certify the
ground state with probability above $0.97$ (dashed line) from $T=12$ on.}
\label{figcred}
\end{figure}

\emph{Certified ground states.---}One desirable property of the swap
is that it records what it repairs. An error swapped out at $s_w$
leaves a photon in the cavity, so the photon, counted at the end of
the anneal \cite{Gleyzes2007,Schuster2007,Blais2021}, tells
whether the run was repaired, thus certifying the ground state. Figure~\ref{figcred} shows the reliability of this certificate,
the probability that the register holds the ground state given the
presence of the photons. At $T=12$ a run with one photon is correct
with probability $0.39$, eight times the $0.05$ of a run with none: at
short $T$ nearly every run acquires an error, so vacuum most likely
means the error still sits in the register, while a photon records
that it was swapped out. However, the one-photon certificate is imperfect:
a swap from $|E^{\rm R}_2,0\rangle$ to $|E^{\rm R}_1,1\rangle$ leaves
one photon but an excited register.

Yet, higher photon numbers certify the ground state with near
certainty. Population that passes $s_2$ diabatically drops at a later
crossing onto a higher ferromagnetic level, $k$ excitations above the
ground level. It returns to the ground level at the $k$-photon
resonance $E^{\rm R}_k-E^{\rm R}_0=k\omega_c$ and leaves $k$ photons
in the cavity (End Matter), because each action of the coupling
$g(a+a^\dagger)O$ takes one excitation from the register and adds one
photon to the cavity. Each photon thus marks a level the register came
down. If the register is still excited when the cavity holds $k$
photons, it must have come down from a level above $k$, which happens
ever more rarely as $k$ grows. The reliability of the $k$-photon
certificate therefore grows with the photon number: runs with three photons are
already correct with probability above $0.97$ at $T=12$
(Fig.~\ref{figcred}) and approach unity much faster than the overall
success probability, as discussed below.

The main cost is the yield of the states with photons, which is modest but
within a practical range: a run returns two photons with probability
$5\%$ and three photons with probability $0.1\%$, averaged over
$T>10$ in Fig.~\ref{figcred}. In a
real quantum simulator, however, the coherence time rather than the
number of runs is the limiting factor: short anneals can be repeated \cite{King2022,King2023,King2025},
while an anneal longer than the coherence time destroys the
ground state. This is the case in Fig.~\ref{figcred}: bringing the
overall success probability to $0.99$
takes several times the annealing time at which runs with three photons
reach it.

\begin{figure}[t]
\includegraphics[width=\columnwidth]{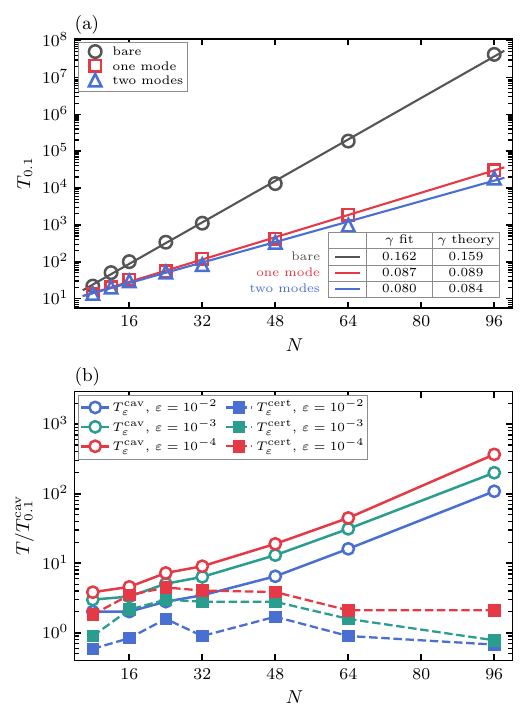}
\caption{How the gain grows with size. (a) Annealing time $T_{0.1}$ versus $N$, each $N$ at its optimal $(\omega_c,\lambda)$,
within $\omega_c=2.25$--$3.5$, $\lambda=0.5$--$0.7$, with
exponential fits $e^{\gamma N}$; the table compares the fitted exponents with
Eq.~\eqref{eq:gammacav}. (b) The advantage of photon certification grows rapidly with $N$.
Times relative to $T^{\rm cav}_{0.1}$, one mode:
$T^{\rm cav}_\varepsilon$, at which
the error of all runs falls below $\varepsilon$ for good, and
$T^{\rm cert}_\varepsilon$, at which runs with at least $k$ photons
($k$ chosen at each $T$) reach error $\varepsilon$ with a yield of at
least $10^{-4}$.}
\label{figscale}
\end{figure}

\emph{Scaling.---}Numerical data show that the bare annealing time to
reach error $0.1$, $T^{\rm bare}_{0.1}$, grows as $e^{0.16N}$, while
with cavity quantum annealing $T^{\rm cav}_{0.1}$ grows as $e^{0.087N}$
[Fig.~\ref{figscale}(a)]. The exponent is almost halved,
$T^{\rm cav}_{0.1}\approx(T^{\rm bare}_{0.1})^{1/2}$, which can be
explained by multiphoton swaps. We consider $\lambda>\lambda^*$, where
the error is limited by the ferromagnetic ladder (Fig.~\ref{figE1}).
For a general $N$, past $s^*$ the paramagnetic level anticrosses a
series of ferromagnetic levels, at $s_k$ with gap $\Delta_k$ and rate
$\Gamma_k=\pi\Delta_k^2/(2v_k)$. Each adiabatic sweep through a crossing drops the
population onto the ferromagnetic level $k-1$, from which a
$(k-1)$-photon swap brings it back to the ground level. A swap with
more photons is a higher-order process with a smaller gap, which the
anneal passes adiabatically only on the lower levels: errors on levels
below a cutoff $k_{\rm c}$ are swapped out, and errors that reach level
$k_{\rm c}$ remain. Thus, the main path that leaves an error crosses
$s^*,s_2,\dots,s_{k_{\rm c}}$ diabatically in succession, and the error survives with
probability $\exp(-T\sum_{k\le k_{\rm c}}\Gamma_k)$. The gaps grow with
$k$, because the paramagnetic level overlaps more with the higher
ferromagnetic levels:
$\Gamma_k/\Gamma_1=(c^2N)^{k-1}/(k-1)!$ with a constant $c$ (see End
Matter). The sum is thus dominated by its largest term, and
$T^{\rm cav}_{0.1}=\ln10/\Gamma_{k_{\rm c}}$. We observe that the cutoff
follows a simple linear law, $k_{\rm c}-1=\nu N$, with the cutoff growth rate
$\nu$ equal to the rate at which the mean photon number of the runs
that end in the ground level grows (End Matter). With
$T^{\rm bare}_{0.1}=\ln10/\Gamma_1\sim e^{\gamma_1N}$, this gives
\begin{equation}
T^{\rm cav}_{0.1}\sim e^{\gamma_{\rm cav}N},\qquad
\gamma_{\rm cav}=\gamma_1-\nu\left(1+\ln\frac{c^2}{\nu}\right).
\label{eq:gammacav}
\end{equation}
The first term is the bare exponent, $\gamma_1=0.159$ from the
spectrum alone, which matches the $0.162$ fitted to the bare dynamics;
the second is what the cavity removes. Eq.~\eqref{eq:gammacav} reproduces the fitted
exponent very well [Fig.~\ref{figscale}(a)].

Eq.~\eqref{eq:gammacav} also points the way to a larger speedup: the
faster the cutoff rises, the more of the exponent is removed. Adding
cavity modes raises $\nu$, because each frequency brings its own swap
resonances. With two modes at distinct frequencies, $\nu$ grows from
$0.035$ to $0.040$, which results in a lower exponent, $0.084$, in
agreement with the fitted $0.080$ [Fig.~\ref{figscale}(a)].

Scaling data also show that photon certification becomes more effective
as $N$ grows: runs with enough photons reach high-fidelity ground states
within a few
$T^{\rm cav}_{0.1}$ at every $N$, whereas without certification this
time grows exponentially with $N$, and so does the gain from
certification [Fig.~\ref{figscale}(b)].

\begin{figure}[t]
\includegraphics[width=\columnwidth]{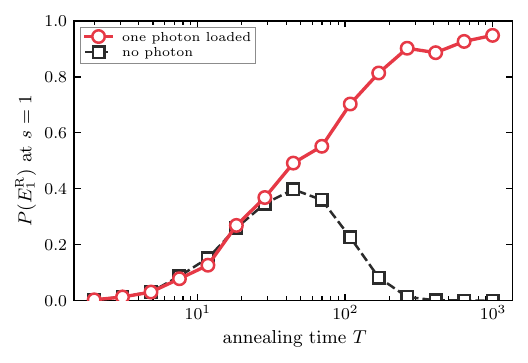}
\caption{The same swap can add an excitation to the register, with one photon loaded. Data for
$N=24$, $\lambda=0.5$. Population $P(E^{\rm R}_1)$ of the first excited
level at $s=1$ versus annealing time, starting from $|E^{\rm R}_0,1\rangle$
(reverse swap) or, for comparison, from $|E^{\rm R}_0,0\rangle$; each
point at its best $\omega_c$.}
\label{fig5}
\end{figure}

\emph{Reverse swap.---}Under the same forward schedule, the
swap can not only remove an excitation from the register but also add
one. With a photon pre-loaded, the system enters the anticrossing at
$s_w$ on the other of the two crossing levels, and an adiabatic sweep
runs the swap the other way, $|E^{\rm R}_0,1\rangle\to|E^{\rm R}_1,0\rangle$:
one photon becomes one collective excitation of the register.
Figure~\ref{fig5} shows the population of $E^{\rm R}_1$ at the end of
the anneal. With one photon loaded, this population grows from $0$
toward $1$ as the annealing time increases. Without the photon
the same protocol reaches $E^{\rm R}_1$ only through uncontrolled
diabatic crossings. Cavity quantum annealing thus opens a direct
route to preparing excited states \cite{LawEberly1996}.

\emph{Outlook.---}We have shown that coupling the annealer to a cavity
mode during the anneal swaps annealing errors into photons. The swap
speeds up the anneal, the photon count certifies the ground state, and
a pre-loaded photon runs the swap in reverse to prepare an excited
state. Looking forward, more cavity modes may accelerate the anneal
further. How far this acceleration goes, and whether it changes the
scaling of the annealing time qualitatively, for example from
exponential to polynomial, are open questions. Applying the mechanism
to other models will require suitable coupling schemes. Superconducting
annealers already couple qubits to microwave resonators
\cite{King2022,Puri2017}, which places the protocol within reach of
current hardware and makes experimental tests the natural next step.
More broadly, cavity quantum annealing offers a new perspective on
quantum computation, in which the cavity, usually a bus or a readout
\cite{Blais2021}, becomes an active part of the computation that
removes errors and records them. This perspective may help design new
quantum error-correction schemes \cite{Young2013,SarovarYoung2013,Matsuura2016,Matsuura2019,Kapit2016,Gertler2021,Sivak2023} and
extend quantum algorithms to processors in which different degrees
of freedom compute together \cite{Liu2026hybrid}.

We thank Matthew Otten and Braden M. Weight for helpful discussions on
cavity quantum electrodynamics. The code for cavity quantum annealing is
available on GitHub at
\url{https://github.com/hao-zhang-quantum/cavity-quantum-annealing}.%

\bibliography{refs}

\onecolumngrid\clearpage%

\section*{End Matter}%
\vspace{2pt}\noindent\includegraphics[width=\textwidth]{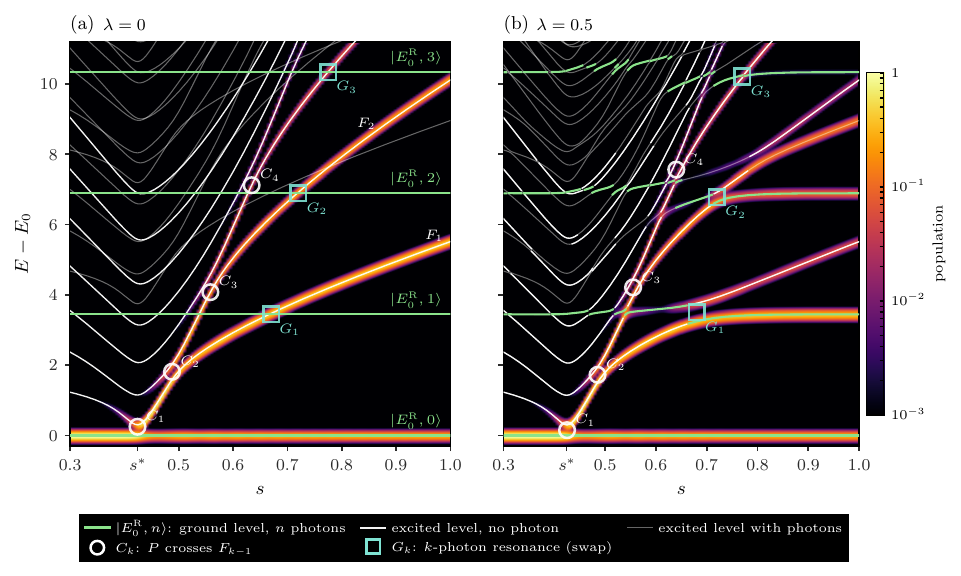}
\refstepcounter{figure}\label{figE1}
\makeatletter\@makecaption{\fnum@figure}{Multiphoton swaps in a fast
anneal, $T=50$, at the parameters of Fig.~\ref{fig1}; color: population
(log scale). (a)~$\lambda=0$: the uncoupled levels
$|E^{\rm R}_k,n\rangle$. $P$ is the paramagnetic level, identified by its
large $\langle O\rangle$, which starts as the ground level and climbs the ferromagnetic ladder after $s^*$, and
$F_k$ is the $k$-th ferromagnetic level. $C_k$ marks the crossing of $P$
with $F_{k-1}$, at $s_k$ in the main text, and $G_k$ the $k$-photon
resonance of $F_k$ with $|E^{\rm R}_0,k\rangle$,
$E^{\rm R}_k-E^{\rm R}_0=k\omega_c$. (b)~$\lambda=0.5$: the exact levels
of register plus cavity, each labeled by the product state
$|E^{\rm R}_k,n\rangle$ onto which it has the largest projection. The
spin--photon coupling $g(a+a^\dagger)O$ opens the crossings between
levels with different photon numbers; at $G_k$ the population on $F_k$ passes to
$|E^{\rm R}_0,k\rangle$, and $G_1$ is the swap at $s_w$.}\makeatother
\vspace{6pt}
\twocolumngrid

\emph{Multiphoton swaps.---}Figure~\ref{fig1} follows one error through
the swap at $s_w$. When the anneal is faster, the error can climb higher
before it is repaired, and swaps with several photons come into play.
Figure~\ref{figE1} shows such an anneal, at the parameters of
Fig.~\ref{fig1} but with $T=50$ instead of $169$, short enough that
the population splits into several branches that can be followed by eye.

Without coupling [Fig.~\ref{figE1}(a)], the population that passes
$s^*$ diabatically stays on the paramagnetic level $P$, which climbs
the ferromagnetic ladder. At each crossing $C_k$ part of it drops onto
the ferromagnetic level $F_{k-1}$, and the rest climbs on. The error
thus spreads over several excited levels and stays there: only $0.27$
of the population ends in the ground level.

With coupling [Fig.~\ref{figE1}(b)], each ferromagnetic level $F_k$
meets the ground level with $k$ photons, $|E^{\rm R}_0,k\rangle$, at
the $k$-photon resonance $G_k$. The coupling turns this crossing into
an avoided one, and the population that follows it leaves the register
in its ground level and $k$ photons in the cavity. The photon number
therefore counts the levels the register came down. At the end of the anneal the
ground level holds $0.26$ of the population with no photon, $0.35$
with one photon, $0.08$ with
two, and $0.002$ with three: $0.70$ in total, against $0.27$ without
coupling.

The shares fall with the photon number because fewer errors climb to
the higher levels [Fig.~\ref{figE1}(a)], and in addition because a
$k$-photon swap is a process of order $k$ in the coupling. The gap at $G_k$ falls from
$0.51$ at $G_1$ to $0.27$ at $G_2$ and $0.07$ at $G_3$, and a fast
anneal passes the smaller gaps diabatically. An error that climbs to a
sufficiently high level therefore misses its swap and is still in the
register at $s=1$. The lowest such level is the cutoff $k_{\rm c}$, the
input to the scaling exponent below.

\emph{Scaling exponent.---}Here we derive the ratio of the error decay
rates $\Gamma_k$ at the successive crossings of the paramagnetic level
with the ferromagnetic ladder [step (ii)], quoted in
the main text, Eq.~\eqref{eq:em_rate}, and from it the
exponent $\gamma_{\rm cav}$ of Eq.~\eqref{eq:gammacav}. Throughout, $N$
is the number of spins, $T$ the annealing time, $F_{\rm spin}$ the
probability that the register ends in its ground level, and $T_{0.1}$
the annealing time at which the error $1-F_{\rm spin}$ reaches $0.1$;
the superscripts bare and cav mark the anneal without and with the
cavity. These rates follow from the spectrum alone; the only input from
the dynamics is the cutoff growth rate $\nu$, the number of levels
the cutoff rises per spin [step (iii)]. Since only the exponent matters, we keep the terms of
$\ln T$ that grow in proportion to $N$ and drop the rest.

(i)~Bare anneal. Without the cavity only one path leaves an
error: the diabatic sweep through the minimum gap $\Delta$ of the
register at $s^*$, which by Eq.~\eqref{eq:lz} happens with probability
$e^{-\Gamma_1T}$. This probability is the error, and it falls to $0.1$
when $e^{-\Gamma_1T}=0.1$, that is at
\begin{equation}
T^{\rm bare}_{0.1}=\frac{\ln10}{\Gamma_1},\qquad
\Gamma_1=\frac{\pi\Delta^2}{2v_1},
\label{eq:em_bare}
\end{equation}
where $\Gamma_1$ is the error decay rate and $v_1$ the slope that the gap
between the two lowest register levels, $E^{\rm R}_1-E^{\rm R}_0$,
approaches on either side of $s^*$. Fitting this gap near $s^*$ to the
hyperbola $(E^{\rm R}_1-E^{\rm R}_0)^2=\Delta^2+v_1^2(s-s^*)^2$ gives
$\Delta$ and $v_1$, and Eq.~\eqref{eq:em_bare} predicts the bare exponent
$\gamma_1$, defined by $T^{\rm bare}_{0.1}\propto e^{\gamma_1N}$, to
within one percent, with no free parameter. This is the starting point,
$\ln T^{\rm bare}_{0.1}=\gamma_1N$ (dropping a constant).

(ii)~Error decay rates along the ferromagnetic ladder. Past $s^*$ the
error sits on the paramagnetic level $P$, which climbs through the
levels of the ferromagnet, $F_0,F_1,\dots$, with $F_0$ the ground level
past $s^*$ (Fig.~\ref{figE1}). The level $P$ crosses $F_{k-1}$ at $s_k$
($k=1,2,\dots$; $s_1=s^*$); this crossing $C_k$ has gap $\Delta_k$
($\Delta_1=\Delta$), relative slope $v_k$, and error decay rate
$\Gamma_k=\pi\Delta_k^2/(2v_k)$. We need the ratio
$\Gamma_k/\Gamma_1=(\Delta_k/\Delta)^2\,v_1/v_k$. In the exact spectrum we observe that the slope ratio $v_1/v_k$ is
around one and barely changes with $N$, so we drop it.

For the gaps we use a semiclassical picture: at large $N$ the
collective spin $\mathbf S=\tfrac12\sum_i\boldsymbol\sigma_i$, of length
$N/2$, behaves as a particle in a double
well~\cite{BapstSemerjian2012,Jorg2010}, with the magnetization
$2S_z/N$ as its coordinate. Its potential has two minima: the paramagnetic well at zero
magnetization, $2S_z/N=0$, and the ferromagnetic well at a
magnetization around one. We treat both wells as harmonic oscillators
of the same width. Then $P$ is the ground state of the paramagnetic well and $F_k$
the $k$-th level of the ferromagnetic well. The gap $\Delta_k$ is
proportional to the overlap of $P$ with $F_{k-1}$. Seen from the
ferromagnetic well, $P$ is the ground state $F_0$ shifted by the
distance between the wells, a coherent state. Its overlap with the
unshifted level $F_m$ is the Poisson amplitude
$\langle F_m|P\rangle=e^{-S/2}S^{m/2}/\sqrt{m!}$~\cite{GerryKnight2023};
here $\sqrt S$ measures the distance between the wells relative to
their zero-point width, and one can show that $\sqrt S=c\sqrt N$ with a
constant $c$. Dividing the overlaps with $F_{k-1}$ and $F_0$, in which
$e^{-S/2}$ cancels, gives the ratio of the gaps:
\begin{equation}
\frac{\Delta_k}{\Delta}=\frac{(c\sqrt N)^{k-1}}{\sqrt{(k-1)!}}.
\label{eq:em_gap}
\end{equation}
Setting $k=2$ in Eq.~\eqref{eq:em_gap} gives $c=\Delta_2/(\Delta\sqrt N)$:
the first two gaps fix the whole ladder. The picture predicts that
$\Delta_2/\Delta$ grows as $\sqrt N$, and the exact spectrum confirms it,
with $c\approx0.31$ at all sizes. With this single constant,
Eq.~\eqref{eq:em_gap} reproduces the exact gaps to within about ten
percent up to the cutoff, $k\le k_{\rm c}$ [step (iii)]. Squaring
Eq.~\eqref{eq:em_gap} gives
\begin{equation}
\frac{\Gamma_k}{\Gamma_1}=\frac{(c^2N)^{k-1}}{(k-1)!}.
\label{eq:em_rate}
\end{equation}

(iii)~Cutoff. The cutoff $k_{\rm c}$ is the lowest ferromagnetic level
whose swap the anneal misses: errors that land on $F_k$ with
$k<k_{\rm c}$ are swapped out, and those that reach $F_{k_{\rm c}}$
remain. An error thus survives only if it passes
$s^*,s_2,\dots,s_{k_{\rm c}}$ diabatically,
\begin{equation}
1-F_{\rm spin}=\exp\Big(-T\sum_{k=1}^{k_{\rm c}}\Gamma_k\Big).
\label{eq:em_survive}
\end{equation}
To use Eq.~\eqref{eq:em_survive} we need the dependence of $k_{\rm c}$ on
$N$. Since the coupling changes the photon number by one, a $k$-photon swap
appears only at $k$-th order in perturbation theory, and its gap shrinks
geometrically as $k$ grows. The time the swap needs thus grows
exponentially with $k$, while the time available, $T_{0.1}$, grows
exponentially with $N$. The photon number of the deepest swap that completes,
$k_{\rm c}-1$, therefore rises linearly with $N$: $k_{\rm c}-1=\nu N$,
with $\nu$ the cutoff growth rate. Numerically,
$\nu=0.035$. We also observe that the mean photon number of the runs
that end in the ground level grows with the same slope. By
Eq.~\eqref{eq:em_rate}, each term of the sum in
Eq.~\eqref{eq:em_survive} then exceeds the previous one by the factor $c^2N/(k-1)\ge c^2/\nu\approx3$, so the last term
dominates and $T^{\rm cav}_{0.1}\approx\ln10/\Gamma_{k_{\rm c}}$. Dividing by
Eq.~\eqref{eq:em_bare} and inserting Eq.~\eqref{eq:em_rate},
\begin{equation}
\ln T^{\rm cav}_{0.1}=\ln T^{\rm bare}_{0.1}-(k_{\rm c}-1)\ln(c^2N)
+\ln\big((k_{\rm c}-1)!\big).
\label{eq:front}
\end{equation}

(iv)~Cavity exponent. Inserting $k_{\rm c}-1=\nu N$ into
Eq.~\eqref{eq:front} and using Stirling's formula, the $N$ inside the
logarithms cancels; with $\ln T^{\rm bare}_{0.1}=\gamma_1N$ this gives
$\ln T^{\rm cav}_{0.1}=\gamma_{\rm cav}N$ with
\begin{equation}
\gamma_{\rm cav}=\gamma_1-\nu\left(1+\ln\frac{c^2}{\nu}\right),%
\label{eq:em_gammacav}
\end{equation}
which is Eq.~\eqref{eq:gammacav} of the main text: the cavity lowers
the bare exponent by $\nu(1+\ln c^2/\nu)$.

\end{document}